\documentclass[pra,preprint,twocolumns,showpacs]{revtex4-1}
\usepackage{graphicx, amsmath}
\begin{document}


\title{Quantum Search with General Nonlinearities}

\author{David A. Meyer}
	\affiliation{\mbox{Department of Mathematics, University of California, San Diego}, \\ La Jolla, CA 92093-0112}
	\email{dmeyer@math.ucsd.edu}
\author{Thomas G. Wong}
	\affiliation{\mbox{Department of Physics, University of California, San Diego}, \\ La Jolla, CA 92093-0354}
	\email{tgw002@physics.ucsd.edu}

\begin{abstract}
	Evolution by the Gross-Pitaevskii equation, which describes Bose-Einstein condensates under certain conditions, solves the unstructured search problem more efficiently than does the Schr\"odinger equation, because it includes a cubic nonlinearity, proportional to $|\psi|^2\psi$.  This is not the only nonlinearity of the form $f(|\psi|^2)\psi$ that arises in effective equations for the evolution of real quantum physical systems, however:  The cubic-quintic nonlinear Schr\"odinger equation describes light propagation in nonlinear Kerr media with defocusing corrections, and the logarithmic nonlinear Schr\"odinger equation describes Bose liquids under certain conditions.  Analysis of computation with such systems yields some surprising results; for example, when time-measurement precision is included in the resource accounting, searching a ``database'' when there is a single correct answer may be easier than searching when there are multiple correct answers.  In each of these cases the nonlinear equation is an effective approximation to a multi-particle Schr\"odinger equation, for search by which Grover's algorithm is optimal.  Thus our results lead to quantum information-theoretic bounds on the physical resources required for these effective nonlinear theories to hold, asymptotically.
\end{abstract}


\maketitle


\section{introduction}

Extensive experimental work has shown that, at least in the familiar regimes of atomic and optical physics, the effect of any fundamental nonlinear generalization of quantum mechanics must be tiny \cite{Weinberg1989, Bollinger1989, Sinha2010}. Nevertheless, there are quantum mechanical systems with multiple interacting particles in which the effective evolution of a single particle is governed by a nonlinear equation. These include Bose-Einstein condensates (BECs) \cite{Bose1924, Einstein1924, Einstein1925}, in which the evolution at low temperatures and densities (so that only two-body contact interactions contribute and the $s$-wave scattering length $a$ is much smaller than the interparticle spacing) is approximately described by a nonlinear Schr\"odinger equation of the Gross-Pitaevskii type \cite{G1961, P1961}:
\[ i \hbar \frac{\partial}{\partial t} \psi(\mathbf{r},t) = \left[ H_0 + \frac{4\pi\hbar^2a}{m} N_0 |\psi(\mathbf{r},t)|^2 \right] \psi(\mathbf{r},t), \]
where $H_0$ includes the kinetic energy and trapping potential, $m$ is the mass of the condensate atom, and $N_0$ is the number of condensate atoms.

\begin{figure}
	\includegraphics{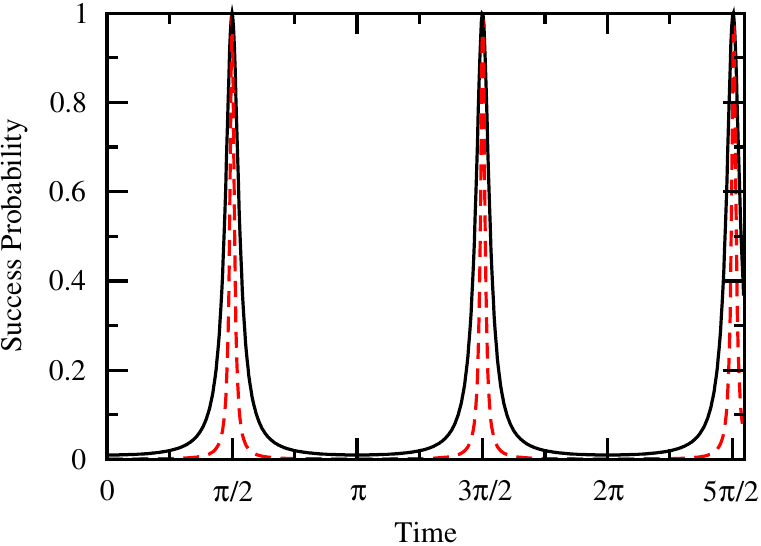}
	\caption{\label{fig:prob_time_critical}Success probability as a function of time for search using the cubic nonlinear Schr\"odinger equation with $k = 1$ marked sites and appropriate parameters in \cite{MeyerWong2013}. The black solid curve is $N = 100$ and the red dashed curve is $N = 1000$, illustrating a constant-runtime solution.}
\end{figure}

In a previous paper \cite{MeyerWong2013}, we quantified the computational advantage that this cubic nonlinear Schr\"odinger equation has in solving the unstructured search problem. To summarize, we search for one of $k$ ``marked'' basis states among $N$ orthonormal basis states $\{|0\rangle, |1\rangle, \dots, |N-1\rangle\}$. Without the nonlinearity, the optimal solution is the continuous-time analogue of Grover's algorithm \cite{Grover1996, Zalka1999, FG1998, Cleve2009}, which runs in time $O(\sqrt{N/k})$. With the nonlinearity, we can search in constant time with appropriate choice of parameters, as shown in FIG.~\ref{fig:prob_time_critical}. This figure also reveals that the success probability spikes suddenly, so increasingly precise time measurement is necessary catch the spike. This requires a certain number of atoms in an atomic clock that utilizes entanglement \cite{GLM2004, BIWH1996}. Jointly optimizing the runtime and number of clock ions, we achieve a resource requirement of $O((N/k)^{1/4})$---a square-root speedup over the linear quantum algorithm.

As explained in \cite{MeyerWong2013}, Grover's algorithm is optimal \cite{Zalka1999}, so there must be additional resources such that the product of the space requirements and the square of the time requirements is lower bounded by $N$. In the case of the Gross-Pitaevskii equation, the additional resource is the condensate atoms, and the bound on the number of them is strongest at $\Omega(N/\log N)$ for the constant-runtime algorithm. Thus we've found a quantum information-theoretic lower bound on the number of condensate atoms needed for the Gross-Pitaevskii equation to be a good asymptotic description of the many-body, linear dynamics.

These two results---a significant, but not unreasonable, square-root speedup in solving the unstructured search problem, and the lower bound on the resources necessary for the Gross-Pitaevskii equation to be valid---suggests it is valuable to quantify the computational advantage that other effective nonlinearities have in solving the unstructured search problem. In particular, we consider nonlinear Schr\"odinger equations of the form
\begin{equation}
	\label{eq:NLSE}
	i \frac{\partial \psi}{\partial t} = \left[ H_0 - g f(|\psi|^2) \right] \psi,
\end{equation}
where $f$ is some real-valued function. The cubic nonlinear Schr\"odinger equation is the case when $f(p) = p$.

A reasonable way to adjust the cubic nonlinearity is to include higher-order terms, such as the quintic term that appears when three-body interactions are included in the description of a BEC \cite{Gammal2000}. Another example of including higher-order terms is the propagation of light in Kerr media \cite{Kerr1877, Kerr1878, Weinberger2008}, whose quantum origins are worked out in \cite{Takatsuji1967}. When a material is subjected to an electric field $E$, its index of refraction $n$ changes:
\[ n(E) = n + \frac{{\rm d}n}{{\rm d}E} E + \frac{1}{2} \frac{{\rm d}^2n}{{\rm d}E^2} E^2 + \dots \]
But from symmetry, many materials require that the index of refraction be an even function. Then the first-order term is zero, leaving
\[ n(E) = n + \frac{1}{2} \frac{d^2n}{dE^2} E^2 + \dots \]
The electric field needn't come from an external source---it can be from the incident light itself. For certain incident light beams, this second-order correction is self-focusing, and it appears in the equation of motion as a cubic term \footnote{Since the intensity is proportional to the square of the electric field, the index of refraction is frequently written as $n(I) = n_0 + n_2I$.}. The cubic self-focusing term, however, is sometimes insufficient to describe the propagation, and a quintic defocusing correction must be included \cite{Smektala2000, Boudebs2003, Zhan2002}. This results in the cubic-quintic nonlinear Schr\"odinger equation
\[ i \frac{{\rm d} \psi_n}{{\rm d}t} = -\gamma N \Delta \psi_n - g \left( |\psi_n|^2 + |\psi_n|^4 \right) \psi_n, \]
which naturally describes a periodic array of $N$ waveguides, where $\gamma$ is a parameter, $\psi_n$ is the amplitude of the electromagnetic wave in each waveguide, and $\Delta$ is the discrete second derivative \cite{CarreteroGonzalez2006}. This equation is of the form of \eqref{eq:NLSE} with $f(p) = p - p^2$.

The above nonlinearities, and indeed general nonlinearities of the form \eqref{eq:NLSE}, do not retain the separability of noninteracting subsystems. That is, in (linear) quantum mechanics, if a physical system consists of two noninteracting subsystems, then its state can be written as the product of the states of the subsystems (\textit{i.e.}, as product state). Nonlinearities, however, generally cause initially uncorrelated subsystems to become correlated. The one exception \cite{BB1976} is the special case when $f(p) = \log(p)$. Then separability is retained, and the nonlinear Schr\"odinger equation \eqref{eq:NLSE} contains a loglinear term:
\[ i \frac{\partial \psi}{\partial t} = \left[ H_0 - g \log(|\psi|^2) \right] \psi. \]
Note that the limit of $\sqrt{x}\log(x)$ as $x$ goes to $0$ is $0$, so the evolution doesn't cause the wavefunction to diverge \footnote{This concern was also addressed in \cite{Avdeenkov2011} by examining the generalized Lagrangian density and effective potential density in \cite{BB1976}.}. Not only is the logarithmic \footnote{Although the nonlinearity is loglinear, the equation is typically referred to as logarithmic. This is different from the cubic and cubic-quintic nonlinearites where the equations are also referred to as cubic and cubic-quintic, respectively.} nonlinear Schr\"odinger equation important for its uniqueness in retaining separability, but it may be suitable for describing Bose liquids, which have higher densities than BECs \cite{Avdeenkov2011}.

In the following section, we formally introduce the unstructured search problem. Then we solve it using the general nonlinear Schr\"odinger equation \eqref{eq:NLSE}, referencing the solution to the cubic nonlinear Schr\"odinger equation from \cite{MeyerWong2013} as we go. Finally, we end with two comprehensive examples of searching with the cubic-quintic and loglinear nonlinearities that were introduced above and give lower bounds for the physical resources needed for them hold.

\section{Setup}

In the unstructured search problem, we look for one correct answer in a ``database'' of $N$ items, of which there are $k$ correct answers. Formally, the system evolves in a $N$-dimensional Hilbert space with computational basis $\{ | 0 \rangle, \dots, | N-1 \rangle\}$. The initial state $| \psi(0) \rangle$ is an equal superposition $| s \rangle$ of all these basis states:
\[ | \psi(0) \rangle = | s \rangle = \frac{1}{\sqrt{N}} \sum_{i=0}^{N-1} | i \rangle. \]
The goal is to evolve the system in such a way that a measurement yields a ``marked'' basis state. There are $k$ marked basis states, of which we only need one, and we call the set of them $M$.

Classically, since there is no structure or ordering to the database elements, one would have to check each item one-by-one until a correct answer is found, which on average takes $O(N/k)$ trials.

Quantum mechanically (\textit{i.e.}, without nonlinearities), we evolve the system according to Schr\"odinger's equation with Hamiltonian
\[ H_0 = -\gamma N | s \rangle \langle s | - \sum_{x \in M} | x \rangle \langle x |, \]
where $\gamma$ is a parameter, inversely proportional to mass. The first term effects a quantum random walk on the complete graph \cite{CG2004}, and the second term is a potential well at the marked sites, causing amplitude to build up there. When $\gamma$ takes a critical value of $\gamma_L = 1/N$, this Hamiltonian is the one that governs  Farhi and Gutmann's \cite{FG1998} ``analog analogue'' of Grover's algorithm (generalized to multiple marked sites); let's call it $H_\text{FG}$. It optimally achieves the search with probability $1$ in time $O(\sqrt{N/k})$, which is a square-root speedup over the classical algorithm.

In the nonlinear regime, we include an additional nonlinear ``self-potential''
\[ V(t) = g \sum_{i=0}^{N-1} f\!\left(\left| \langle i | \psi \rangle \right|^2\right) | i \rangle \langle i |, \]
which we subtract from $H_0$ so that the system evolves according to the generalized nonlinear Schr\"odinger equation \eqref{eq:NLSE}. As proved in \cite{MeyerWong2013}, $g$ must be greater than zero for the cubic nonlinear algorithm to perform better since, heuristically, it causes the self-potential to act as an additional potential well, therefore attracting more probability and speeding up the search. So we require $g > 0$ for our general nonlinearity as well, and greater $g$ should result in a faster algorithm.

As the system evolves, it remains in the two-dimensional subspace spanned by orthonormal vectors
\[ \frac{1}{\sqrt{k}} \sum_{i \in M} | i \rangle \quad \text{and} \quad \frac{1}{\sqrt{N-k}} \sum_{i \notin M} | i \rangle, \]
so we can write $| \psi(t) \rangle$ as a linear combination of them:
\[ | \psi(t) \rangle = \alpha(t) \frac{1}{\sqrt{k}} \sum_{x \in M} | x \rangle + \beta(t) \frac{1}{\sqrt{N-k}} \sum_{x \notin M} | x \rangle. \]
Then the probability of measuring the system in basis state $| i \rangle$ is
\[ | \langle i | \psi \rangle |^2 = \begin{cases}
	\frac{|\alpha|^2}{k}, & i \in M \\
	\frac{|\beta|^2}{N-k}, & i \not\in M \\
\end{cases}. \]
Let's define
\[ f_\alpha = f\!\left( \frac{|\alpha|^2}{k} \right), \quad \text{and} \quad f_\beta = f\!\left( \frac{|\beta|^2}{N-k} \right). \]
Then the nonlinear Schr\"odinger equation \eqref{eq:NLSE} is written in the two-dimensional subspace as
\begin{equation}
	\label{eq:NLSEmatrix}
	\frac{{\rm d}}{{\rm d}t} \begin{bmatrix} \alpha \\ \beta \end{bmatrix} = i \begin{bmatrix}
	\gamma k + 1 + g f_\alpha & \gamma \sqrt{k} \sqrt{N-k} \\
	\gamma \sqrt{k} \sqrt{N-k} & \gamma(N-k) + g f_\beta \end{bmatrix} \begin{bmatrix} \alpha \\ \beta \end{bmatrix}.
\end{equation}

\section{Critical Gamma}

\begin{figure}
	\includegraphics[width=3in]{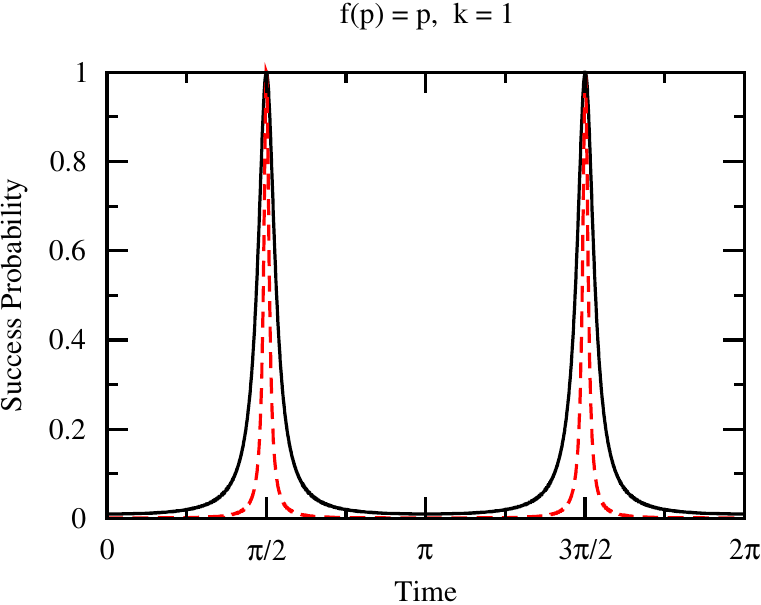}
	\includegraphics[width=3in]{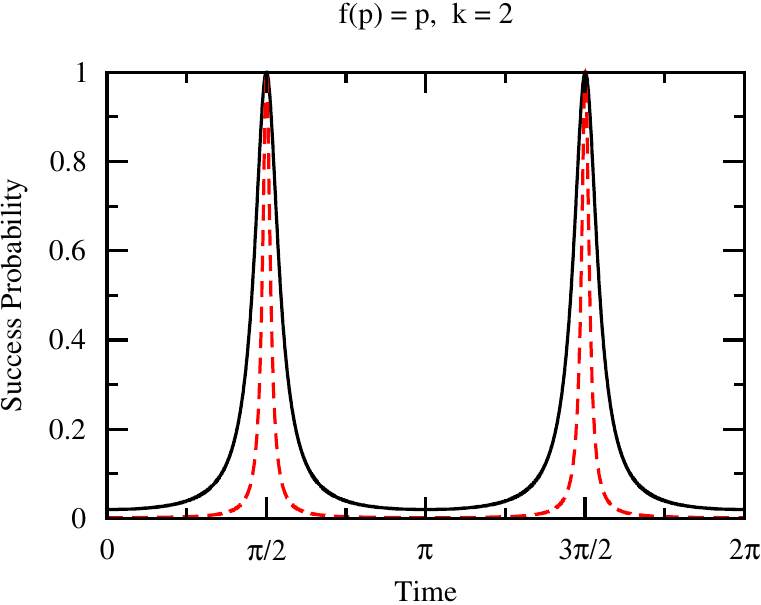}
	\includegraphics[width=3in]{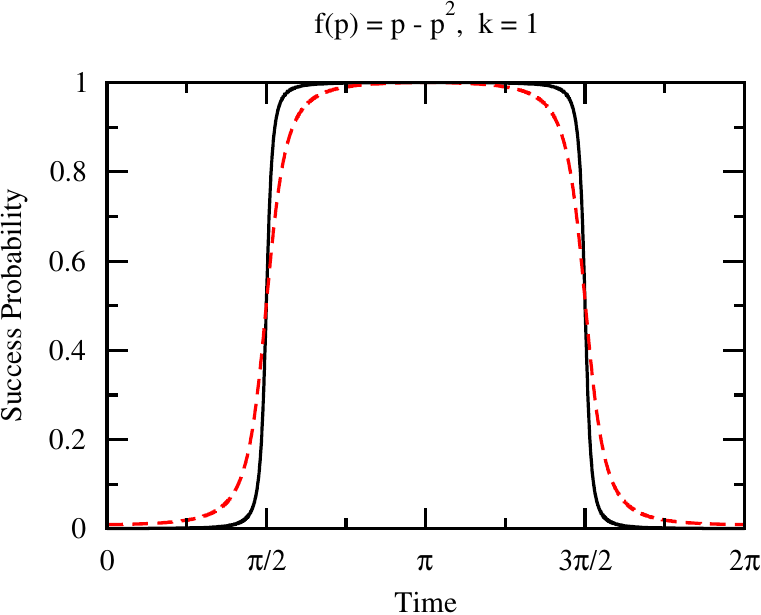}
	\includegraphics[width=3in]{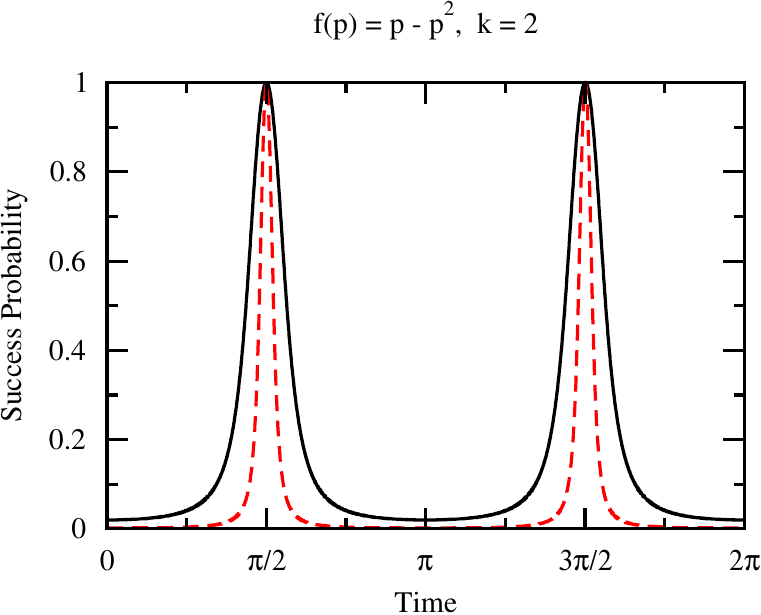}
	\includegraphics[width=3in]{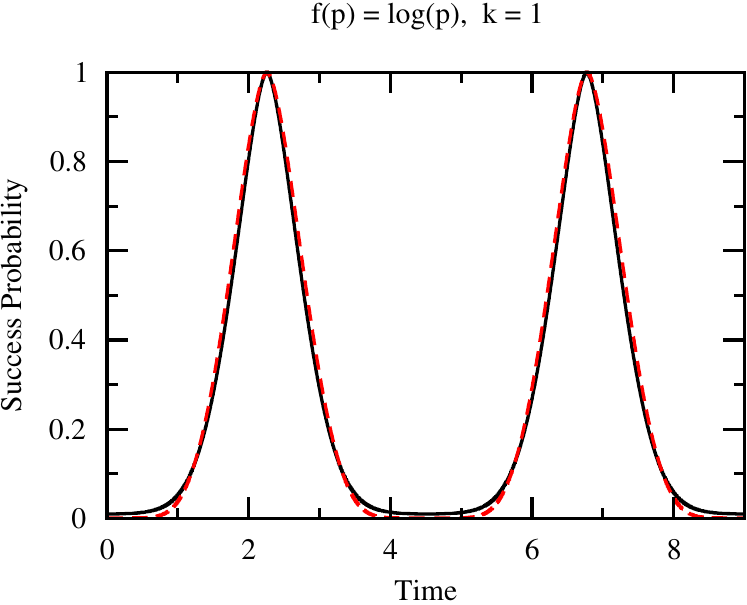}
	\includegraphics[width=3in]{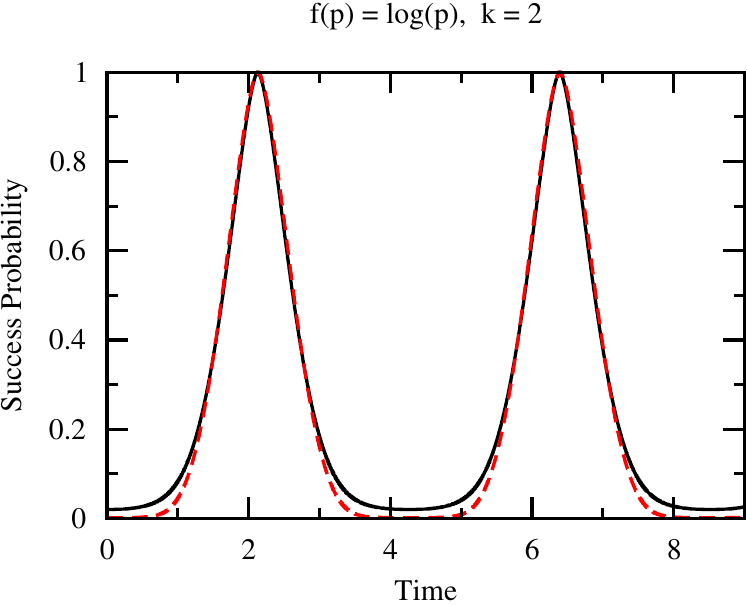}
	\caption{\label{fig:prob_time_mashup}Success probability as a function of time for search using the cubic, cubic-quintic, and loglinear Schr\"odinger equation with $k = 1$ and $k = 2$ marked sites and $\gamma$ at its critical value given by \eqref{eq:gammac}. The black solid curve is $N = 100$ and the red dashed curve is $N = 1000$. The nonlinearity coefficient $g$ scales as $O(N)$, $O(N)$, and $O(\sqrt{N}/\log N)$ for the respective nonlinearities so as to make the runtime constant.}
\end{figure}

We can also write $H(t) = H_0 - V(t)$ in terms of $f_\alpha$ and $f_\beta$:
\begin{align*}
	H &= -\gamma N | s \rangle \langle s | - \sum_{i \in M} | i \rangle \langle i | - g f_\alpha \sum_{i \in M} | i \rangle \langle i | - g f_\beta \sum_{i \notin M} | i \rangle \langle i | \\
	  &= -\gamma N | s \rangle \langle s | - \left( 1 + g f_\alpha - g f_\beta \right) \sum_{i \in M} | i \rangle \langle i | - g f_\beta \sum_{i = 0}^{N-1} | i \rangle \langle i |.
\end{align*}
The last term is a multiple of the identity matrix, which simply redefines the zero of energy (or contributes an overall, non-observable phase), so we can drop it. From our previous work on the cubic nonlinear Schr\"odinger equation \cite{MeyerWong2013}, the critical $\gamma$ causes the nonlinear system to follow the same evolution as the linear, optimal algorithm, but with rescaled time. That is, we choose
\begin{equation}
	\label{eq:gammac}
	\gamma_c = \gamma_L \left[ 1 + g \left( f_\alpha - f_\beta \right) \right] = \frac{1}{N} \left[ 1 + g \left( f_\alpha - f_\beta \right) \right],
\end{equation}
which is time-dependent, so that
\[ H = \left( 1 + g f_\alpha - g f_\beta \right) \left( -\gamma_L N | s \rangle \langle s | - \sum_{i \in M} | i \rangle \langle i | \right) = \left( 1 + g f_\alpha - g f_\beta \right) H_\text{FG}. \]
So the system evolves according to Farhi and Gutmann's Hamiltonian, but with continuously rescaled time. Thus we have the critical $\gamma$ \eqref{eq:gammac} for general nonlinearities of the form \eqref{eq:NLSE}. Note that for the cubic nonlinearity, $f(p) = p$, so if we define $G = g/[k(N-k)]$ and $\delta = (N-k)|\alpha|^2 - k|\beta|^2$, then we get the familiar result $(1+G\delta)/N$ from \cite{MeyerWong2013}. Additionally, the critical $\gamma$ \eqref{eq:gammac} causes the eigenvectors of $H$ to be proportional to $|s\rangle \pm |w\rangle$. As explained in Section 3 of \cite{MeyerWong2013}, this causes the success probability to reach a value of $1$. This is shown in FIG.~\ref{fig:prob_time_mashup} for the cubic, cubic-quintic, and loglinear nonlinearities. A couple of observations are noteworthy. First, the cubic-quintic nonlinearity with one marked site has a wide peak in success probability, but with multiple marked sites, it has a narrow spike. Catching a narrow spike is more difficult than the wide peak, so searching with one marked site is ``easier'' than searching with multiple marked sites. This is counterintuitive, and it will be explicitly proven later. Second, for the loglinear nonlinearity, the success probability has a constant width. For the rest of the paper, we choose $\gamma = \gamma_c$ as defined in \eqref{eq:gammac}.

\section{Runtime}

To derive the runtime of the algorithm, we follow the same procedure given in \cite{MeyerWong2013}, generalized for \eqref{eq:NLSE}. We begin by expliciting writing out \eqref{eq:NLSEmatrix}, which yields two coupled, first-order ordinary differential equations:
\begin{subequations}
	\begin{equation} \label{eq:dadt} \frac{{\rm d}\alpha}{{\rm d}t} = i \left\{ \left[ \gamma k + 1 + g f_\alpha \right] \alpha + \gamma \sqrt{k} \sqrt{N-k} \beta \right\} \end{equation}
	\begin{equation} \label{eq:dbdt} \frac{{\rm d}\beta}{{\rm d}t} = i \left\{ \gamma \sqrt{k} \sqrt{N-k} \alpha + \left[ \gamma (N-k) + g f_\beta \right] \beta \right\}. \end{equation}
\end{subequations}
We define three real variables $x(t)$, $y(t)$, and $z(t)$ such that
\begin{subequations}
	\begin{equation} \label{eq:x} x = |\alpha|^2 \end{equation}
	\begin{equation} \label{eq:yz} y + iz = \alpha \beta^*. \end{equation}
\end{subequations}
Note that $x(t)$ is the success probability, which we want to find. To do this, we want to decouple \eqref{eq:dadt} and \eqref{eq:dbdt} for a single differential equation in terms of $x(t)$ alone and then solve it. We begin by differentiating \eqref{eq:x} by utilizing \eqref{eq:dadt}:
\[ \frac{{\rm d}x}{{\rm d}t} = \frac{{\rm d}|\alpha|^2}{{\rm d}t} = \alpha \frac{{\rm d}\alpha^*}{{\rm d}t} + \frac{{\rm d}\alpha}{{\rm d}t} \alpha^* = 2 \gamma \sqrt{k} \sqrt{N-k} z. \]
Solving for $z$,
\begin{equation}
	\label{eq:z}
	z = \frac{1}{2 \gamma \sqrt{k} \sqrt{N-k}} \frac{{\rm d}x}{{\rm d}t}.
\end{equation}
Note that the critical $\gamma$ depends on $x$:
\[ \gamma_c = \frac{1}{N} \left\{ 1 + g \left[ f\!\left( \frac{x}{k} \right) - f\!\left( \frac{1-x}{N-k} \right) \right] \right\}, \]
so we can use \eqref{eq:z} to eliminate $z$ in favor of $x$ and ${\rm d}x/{\rm d}t$. We can also find an expression for eliminating ${\rm d}z/{\rm d}t$ by differentiating this, but note that $\gamma = \gamma_c$ depends on time. Its derivative is
\[ \frac{d\gamma_c}{{\rm d}t} = \frac{g}{N} \left[ \frac{1}{k} f'_\alpha + \frac{1}{N-k} f'_\beta \right] \frac{{\rm d}x}{{\rm d}t}, \]
where we've defined in analogy to $f_\alpha$ and $f_\beta$,
\[ f'_\alpha = \left. \frac{{\rm d}f(p)}{{\rm d}p} \right|_{p = \frac{|\alpha|^2}{k} = \frac{x}{k}} \quad \text{and} \quad f'_\beta = \left. \frac{{\rm d}f(p)}{{\rm d}p} \right|_{p = \frac{|\beta|^2}{N-k} = \frac{1-x}{N-k}}. \]
Then the derivative of \eqref{eq:z} is
\begin{equation}
	\label{eq:dzdt}
	\frac{{\rm d}z}{{\rm d}t} = \frac{1}{2 \sqrt{k} \sqrt{N-k}} \left\{ \frac{-1}{\gamma^2} \frac{g}{N} \left[ \frac{1}{k} f'\left( \frac{x}{k} \right) + \frac{1}{N-k} f'\left( \frac{1-x}{N-k} \right) \right] \left( \frac{{\rm d}x}{{\rm d}t} \right)^2 + \frac{1}{\gamma} \frac{{\rm d}^2x}{{\rm d}t^2} \right\}.
\end{equation}
So now we can eliminate ${\rm d}z/{\rm d}t$ in favor of $x$, ${\rm d}x/{\rm d}t$, and ${\rm d}^2x/{\rm d}t^2$.

Now let's differentiate \eqref{eq:yz} by utilizing \eqref{eq:dadt} and \eqref{eq:dbdt}, which yields
\[ \frac{{\rm d}}{{\rm d}t} (y + iz) = \frac{{\rm d}(\alpha\beta^*)}{{\rm d}t} = \frac{{\rm d}\alpha}{{\rm d}t} \beta^* + \alpha \frac{{\rm d}\beta^*}{{\rm d}t} = -2 \gamma k z + i \left\{ 2 \gamma k y + \gamma\sqrt{k}\sqrt{N-k}(1-2x) \right\}, \]
where we've used $\gamma = \gamma_c$ to calculate the $2 \gamma k$ coefficients. Matching the real and imaginary parts, we get:
\[ \frac{{\rm d}y}{{\rm d}t} = -2 \gamma k z \]
\[ \frac{{\rm d}z}{{\rm d}t} = 2 \gamma k y + \gamma \sqrt{k} \sqrt{N-k} (1-2x). \]
In the first equation, we can eliminate $z$ using \eqref{eq:z}, which yields
\[ \frac{{\rm d}y}{{\rm d}z} = -\sqrt{\frac{k}{N-k}} \frac{{\rm d}x}{{\rm d}t}. \]
This integrates to
\[ y = -\sqrt{\frac{k}{N-k}} x + \sqrt{\frac{k}{N-k}} = -\sqrt{\frac{k}{N-k}} (x - 1) , \]
where the constant of integration was found using $y(0) = \sqrt{k(N-k)}/N$ and $x(0) = k/N$. Using this to eliminate $y$ in the second equation and simplifying,
\[ \frac{{\rm d}z}{{\rm d}t} = \gamma \sqrt{\frac{k}{N-k}} \left( N - 2Nx + k \right). \]
Eliminating ${\rm d}z/{\rm d}t$ using \eqref{eq:dzdt} and simplifying, we get
\[ \frac{{\rm d}^2x}{{\rm d}t^2} = \frac{1}{\gamma} \frac{g}{N} \left[ \frac{1}{k} f'_\alpha + \frac{1}{N-k} f'_\beta \right] \left( \frac{{\rm d}x}{{\rm d}t} \right)^2 + 2 \gamma^2 k \left( N - 2Nx + k \right), \]
which is entirely in terms of $x$ and its derivatives. Plugging in for $\gamma = \gamma_c$,
\begin{equation}
	\label{eq:d2xdt2}
	\frac{{\rm d}^2x}{{\rm d}t^2} = \frac{N}{1 + g(f_\alpha- f_\beta)} \frac{g}{N} \left[ \frac{1}{k} f'_\alpha + \frac{1}{N-k} f'_\beta \right] \left( \frac{{\rm d}x}{{\rm d}t} \right)^2 + 2 \left( \frac{1+g(f_\alpha-f_\beta)}{N} \right)^2 k \left( N - 2Nx + k \right).
\end{equation}
So we've decoupled \eqref{eq:dadt} and \eqref{eq:dbdt}, yielding a second-order ordinary differential equation for $x$. To solve it, let $h(x) = ({\rm d}x/{\rm d}t)^2$ so that ${\rm d}h/{\rm d}x = 2 {\rm d}^2x/{\rm d}t^2$. Then we get a first-order ordinary differential equation for $h(x)$:
\[ \frac{1}{2} \frac{{\rm d}h}{{\rm d}t} = \frac{N}{1 + g(f_\alpha- f_\beta)} \frac{g}{N} \left[ \frac{1}{k} f'_\alpha + \frac{1}{N-k} f'_\beta \right] h + 2 \left( \frac{1+g(f_\alpha-f_\beta)}{N} \right)^2 k \left( N - 2Nx + k \right). \]
Solving this with the initial condition $h(x=k/N) = 0$, we get
\[ h(x) = \frac{4 k (x-1) (k-Nx) \left[1 + g \left( f_\alpha - f_\beta \right) \right]^2}{N^2}. \]
Taking the square root and noting that ${\rm d}x/{\rm d}t = \pm \sqrt{h(x)}$, 
\begin{equation}
	\label{eq:dxdt}
	\frac{{\rm d}x}{{\rm d}t} = \pm \sqrt{ \frac{4 k (x-1) (k-Nx) \left[1 + g \left( f_\alpha - f_\beta \right) \right]^2}{N^2} } .
\end{equation}
We can solve this using separation of variables and integrating from $t = 0$ to $t$ and $x = k/N$ to $x$, which yields
\begin{equation}
	\label{eq:time}
	t = \frac{N}{2\sqrt{k}} \int_{x_0 = k/N}^{x} \frac{1}{1 + g(f_\alpha - f_\beta)} \sqrt{\frac{1}{(1-x)(Nx-k)}} {\rm d}x.
\end{equation}
This integral depends on the form of $f(p)$. If it is analytically integrable, we get an expression for $t(x)$, which we invert for $x(t)$. For example, for the cubic nonlinearity, $f(p) = p$. Then $f_\alpha - f_\beta = (Nx-k)/(k(N-k))$, and \eqref{eq:time} can be integrated to yield
\begin{equation}
	\label{eq:cubictime}
	t = -\sqrt{\frac{Nk}{k+g}} \left\{ \tan^{-1}\left[\frac{\sqrt{Nk} \sqrt{1-x}}{\sqrt{k+g} \sqrt{N x-k}}\right] - \frac{\pi}{2} \right\},
\end{equation}
which can be solved for a success probability of
\[ x(t) = \frac{N + (k+g) \tan^2 \left[ \frac{\pi}{2} - \sqrt{\frac{k+g}{N}} t \right]}{N + \frac{N}{k} (k+g) \tan^2 \left[ \frac{\pi}{2} - \sqrt{\frac{k+g}{N}} t \right]} . \]
This reaches a value of $1$ at a runtime of
\begin{equation}
	\label{eq:cubicruntime}
	t_* = \frac{1}{\sqrt{k+g}} \frac{\pi \sqrt{N}}{2},
\end{equation}
and the success probability is periodic with period $2t_*$. These results agree with \cite{MeyerWong2013}.

Returning to the general nonlinearity, if we are only interested in the runtime $t_*$ and not the entire evolution of the success probability, then we can instead integrate from $x = k/N$ to $1$:
\begin{equation}
	\label{eq:runtime}
	t_* = \frac{N}{2\sqrt{k}} \int_{x_0 = k/N}^{x_* = 1} \frac{1}{1 + g(f_\alpha - f_\beta)} \sqrt{\frac{1}{(1-x)(Nx-k)}} {\rm d}x.
\end{equation}
Evaluating this for the cubic nonlinearity yields \eqref{eq:cubicruntime}, as expected.

\section{Time-Measurement Precision}

As shown in FIGs.~\ref{fig:prob_time_critical} and \ref{fig:prob_time_mashup}, the spike in success probability may be narrow. To quantify it, let's find the width of the peak at height $1 - \epsilon$.

If we can explicitly integrate \eqref{eq:time}, then we can use the result to find the width in success probability. For example, the cubic nonlinearity yielded \eqref{eq:cubictime}, which we use to find the time at which the success probability reaches a height of $1 - \epsilon$. Then the width of the peak at this height is
\[ \Delta t = 2 \sqrt{\frac{N}{k+g}} \tan^{-1}\left[\frac{\sqrt{Nk} \sqrt{\epsilon}}{\sqrt{k+g} \sqrt{N(1-\epsilon)-k}}\right]. \]
We are interested in how this time-measurement precision scales with $N$, but the inverse tangent makes it difficult to see. Instead, Taylor's theorem can be used to show that it suffices to keep the first term in the Taylor series:
\[ \Delta t^{(0)} = \frac{2Nk}{k+g} \sqrt{\frac{\epsilon}{k(N-k)}} + O(\epsilon^{3/2}). \]
If we define $G = g/[k(N-k)]$, this agrees with our result from \cite{MeyerWong2013}.

For a general nonlinearity, we can find the leading-order formula for the time-measurement precision $\Delta t^{(0)}$ by Taylor expanding the success probability around $x = 1$, which is a maximum so its first-derivative is zero, and using \eqref{eq:d2xdt2} for the second derivative:
\begin{align*}
	x(t) 
		&= x(t_*) + x'(t_*) (t-t_*) + \frac{x''(t_*)}{2} (t-t_*)^2 + ... \\
		&\approx 1 + 0 - \left( \frac{1+g(f_\alpha|_{x=1} - f_\beta|_{x=1})}{N} \right)^2 k(N-k) (t-t_*)^2.
\end{align*}
This reaches a height of $1 - \epsilon$ at times
\[ t \approx t_* \pm \sqrt{\left( \frac{N}{1+g(f_\alpha|_{x=1} - f_\beta|_{x=1})} \right)^2 \frac{\epsilon}{k(N-k)}}. \]
So, the leading-order width of the peak is
\begin{equation}
	\label{eq:width}
	\Delta t^{(0)} = \frac{2N}{1+g(f_\alpha|_{x=1} - f_\beta|_{x=1})} \sqrt{\frac{\epsilon}{k(N-k)}}.
\end{equation}
For the cubic nonlinearity, $f_\alpha|_{x=1} - f_\beta|_{x=1} = 1/k$, so we get
\[ \Delta t^{(0)} = \frac{2N}{1+g/k} \sqrt{\frac{\epsilon}{k(N-k)}}, \]
which agrees with our previous result.

To attain this level of time-measurement precision, say we use an atomic clock with $N_\text{clock}$ entangled ions. Then the time-measurement precision goes as $O(1/N_\text{clock})$ \cite{GLM2004, BIWH1996}. So the number of atomic clock ions we need is inversely proportional to the required time-measurement precision. This, plus the $\log N$ qubits needed to encode the $N$-dimensional Hilbert space, gives the ``space'' requirement of our algorithm. The product of ``space'' and time, which preserves the time-space tradeoff inherent in na\"ive parallelization, gives the total resource requirement.

Now that we have formulas for the runtime \eqref{eq:runtime} and time-measurement precision \eqref{eq:width} for a general nonlinearity of the form \eqref{eq:NLSE}, let's calculate them for the specific examples of the cubic-quintic and loglinear nonlinearities. But for comparison's sake, let's first review the results for the cubic nonlinearity.

\section{Cubic Nonlinearity}

The cubic nonlinear Schr\"odinger equation has the form \eqref{eq:NLSE} with $f(p) = p$. From before, we found
\[ t_* = \frac{1}{\sqrt{k+g}} \frac{\pi \sqrt{N}}{2} \]
and
\[ \Delta t^{(0)} = \frac{2N}{1+g/k} \sqrt{\frac{\epsilon}{k(N-k)}}, \]
both of which agree with \cite{MeyerWong2013}. If $g = O(N^\kappa)$ and $k = O(N^\lambda)$ (with $0 \le \lambda \le 1$), then these become
\[ t_* = \begin{cases}
	O\left( N^{-\kappa/2 + 1/2} \right), & \kappa \ge \lambda \\
	O\left( N^{-\lambda/2 + 1/2} \right), & \kappa < \lambda
\end{cases} \]
and
\[ \Delta t^{(0)} = \begin{cases}
	O\left( N^{-\kappa + \lambda/2 + 1/2} \right), & \kappa \ge \lambda \\
	O\left( N^{-\lambda/2 + 1/2} \right), & \kappa < \lambda
\end{cases}. \]
To achieve this level of time-measurement precision, the number of ions in an atomic clock that utilizes entanglement must scale as the reciprocal of the precision \cite{GLM2004, BIWH1996}. Including the $\log N$ qubits to encode the $N$-dimensional Hilbert space, the total ``space'' requirement $S$ scales as
\[ S = \begin{cases}
	O\left( N^{\kappa - \lambda/2 - 1/2} \right), & \kappa \ge \lambda/2 + 1/2 \\
	O\left( \log N \right), & \kappa < \lambda/2 + 1/2 \\
\end{cases}. \]
Then the total resource requirement is
\[ ST = \begin{cases} 
	O\left( N^{\kappa/2 - \lambda/2}  \right), & \kappa \ge \lambda/2 + 1/2 \\
	O\left( N^{-\kappa/2 + 1/2} \log N \right), & \kappa \ge \lambda, \kappa < \lambda/2 + 1/2 \\
	O\left( N^{-\lambda/2 + 1/2} \log N  \right), & \kappa < \lambda
\end{cases} \]
This takes a minimum value of $ST = N^{-\lambda/4+1/4} \log N = (N/k)^{1/4} \log N$ when $\kappa = \lambda/2 + 1/2$, and it makes the width $\Delta t^{(0)}$ constant.

Of course, the cubic nonlinear Schr\"odinger equation, or Gross-Pitaevskii equation, is an effective nonlinear theory that only approximates the linear evolution of the multiparticle Schr\"odinger equation describing Bose-Einstein condensates. As worked out in \cite{MeyerWong2013} for the case of a single marked vertex, and generalized here to multiple marked vertices, since Grover's algorithm is optimal \cite{Zalka1999} for (linear) quantum computation, the number of condensate atoms $N_0$ must be included in the resource accounting such that the product of the space requirements and the square of the time requirements is lower bounded by $N$. That is, since there are $N_0$ oracles, each responding to a $\log N$ bit query,
\[ ST^2 = \begin{cases} 
	O\left( N^{-\lambda/2 + 1/2} + N^{-\kappa + 1} N_0 \log N  \right), & \kappa \ge \lambda/2 + 1/2 \\
	O\left( N^{-\kappa + 1} N_0 \log N \right), & \kappa \ge \lambda, \kappa < \lambda/2 + 1/2 \\
	O\left( N^{-\lambda + 1} N_0 \log N  \right), & \kappa < \lambda
\end{cases} = \Omega(N). \]
Then
\[ N_0 = \begin{cases}
	\Omega\left( \frac{N^\kappa}{\log N} \right), & \kappa \ge \lambda \\
	\Omega\left( \frac{N^\lambda}{\log N} \right), & \kappa < \lambda \\
\end{cases}. \]
In the first region, this bound is maximized when $\kappa = 1$, corresponding to the constant-runtime solution and beyond which it doesn't make sense to increase $\kappa$. In the second region, it is maximized when $\lambda = 1$, \textit{i.e.}, the number of marked vertices scales with $N$. In both of these cases, the bound takes its strongest value:
\[ N_0 = \Omega\left( \frac{N}{\log N} \right). \]
As expressed in \cite{MeyerWong2013}, to the best of our knowledge, this is the first bound on the number of condensate atoms needed for the Gross-Pitaevskii to be a good approximation of the linear, multiparticle dynamics.

\section{Cubic-Quintic Nonlinearity}

The cubic-quintic nonlinear Schr\"odinger equation has the form \eqref{eq:NLSE} with $f(p) = p - p^2$. Then
\[ f_\alpha - f_\beta = \frac{-N(N-2k)x^2 + k(N^2-kN-2k)x - k^2(N-k-1)}{k^2(N-k)^2}. \]
Plugging this into \eqref{eq:runtime}, the runtime is given by an integral of the form
\[ t_* = \frac{Nk^2(N-k)^2}{2\sqrt{k}} \int_{x_0 = k/N}^{x_* = 1} \frac{1}{ax^2 + bx + c} \sqrt{\frac{1}{(1-x)(Nx-k)}} {\rm d}x, \]
where
\[ a = -g N (N-2k) \]
\[ b = gk(N^2-kN-2k) \]
\[ c = -gk^2(N-k-1) + k^2(N-k)^2. \]
This is analytically integrable, and the solution is
\[ t_* = \frac{\pi}{2} \frac{N k^2(N-k)^2}{2\sqrt{k}} \frac{\sqrt{2}}{\sqrt{\Sigma } \sqrt{\Delta }} \left[ \frac{2 a+b+\sqrt{\Delta}}{\sqrt{\xi +\sqrt{\Delta} (k-N)} } + \frac{-2 a-b+\sqrt{\Delta}}{\sqrt{\xi -\sqrt{\Delta}(k-N)}} \right], \]
where
\[ \Delta = b^2-4a c \]
\[ \Sigma = a+b+c \]
\[ \xi = 2ak + 2cN + b(k+N). \]
It is rather tedious to find the scaling of this runtime when $g = O(N^\kappa)$ and $k = O(N^\lambda)$ (with $0 \le \lambda \le 1$), so the details are in Appendix \ref{appendix:cubicquintic}. The result is
\[ t_* = \begin{cases}
	O\left( N^{-\kappa/2 + 1/2} \right), & \lambda \le \kappa \\
	O\left( N^{-\lambda/2 + 1/2} \right), & \lambda > \kappa \\
\end{cases}, \]
which is the same runtime order as search with the cubic nonlinearity.

For the time-measurement precision, note that $f_\alpha|_{x=1} - f_\beta|_{x=1} = (k-1)/k^2$. Plugging this into \eqref{eq:width}, the width of the spike in success probability at height $1 - \epsilon$ is
\[ \Delta t^{(0)} = \frac{2N}{1+g(k-1)/k^2} \sqrt{\frac{1}{k(N-k)} \epsilon}. \]
When $k = 1$, the $g$ term disappears. So varying $g$, while changing the runtime, doesn't affect the width. When $k \ne 1$, it is
\[ \Delta t^{(0)} = \frac{2N}{1+g/k} \sqrt{\frac{1}{k(N-k)} \epsilon}, \]
which is the same as the cubic nonlinearity's formula. Putting these together and letting $g = O(N^\kappa)$ and $k = O(N^\lambda)$ (with $0 \le \lambda \le 1$), we get
\[ \Delta t^{(0)} = \begin{cases}
	O\left( N^{1/2} \right), & \lambda = 0 \\
	O\left( N^{-\kappa + \lambda/2 + 1/2} \right), & \lambda \ne 0, \lambda \le \kappa \\
	O\left( N^{-\lambda/2 + 1/2} \right), & \lambda \ne 0, \lambda > \kappa
\end{cases}. \]

So the runtime of search with the cubic-quintic nonlinearity scales identically to search with the cubic nonlinearity. Furthermore, when there are multiple marked sites, the time-measurement precision also scales the same. But when there is a single marked site, the time-measurement precision scales as $N^{1/2}$, which is the same as Farhi and Gutmann's linear algorithm \cite{FG1998}. This distinction between single and multiple marked sites is evident in FIG.~\ref{fig:prob_time_mashup}. So for a single marked site, all the speedup that comes from the nonlinearity can be utilized without the expense of increasing the time-measurement precision. Thus search with the cubic-quintic nonlinearity is able to achieve a jointly-optimized runtime and time-measurement precision of $O(1)$.

As explained in \cite{MeyerWong2013}, Grover's algorithm is optimal \cite{Zalka1999}, so there must be additional resources such that the product of the space requirements and the square of the time requirements is lower bounded by $N$. For the cubic-quintic nonlinearity, say the physical system is a periodic array of $N$ waveguides, each long enough that the electromagnetic wave propagating through it performs the calculation. So the length of the waveguide would scale with the runtime $t_*$. Keeping the cross sectional area of the waveguide constant, the number of atoms in a waveguide would also go as $t_*$. Since we have $N$ waveguides, the number of atoms would go as $Nt_*$. If the runtime is constant, then the number of atoms goes as $\Omega(N)$, satisfying the optimality proof's lower bound.

The amount of energy, or number of photons, can also be included in the resource accounting. Say a waveguide needs $P$ photons in the incident beam for it to behave like Kerr media with quintic corrections. Then we would need $PN$ photons for the whole array. But it's reasonable to say $P$ is constant, so this would scale as $N$, again satisfying the optimality proof's lower bound.

We would also need charge to create an electric field at the marked waveguides. Say this takes a constant number of resources. There are $k$ marked waveguides, so the resources for this would scale as $k$. While this may scale less than $N$, the other physical resources already satisfy the optimality proof's lower bound.

These resources may seem excessive, but if they scale linearly with $N$, then it is no different than any other database that requires the $N$ items in the database to be physically written somewhere.

For other physical systems that are effectively described by the cubic-quintic nonlinear Schr\"odinger equation, such as Bose-Einstein condensates with a higher-order corrections \cite{Gammal2000}, the additional resource to the runtime and time-measurement precision is some number of particles $N_0$, each of which responds to a $\log N$ bit query. As proved above, when there are multiple marked vertices, the cubic-quintic nonlinearity solves the unstructured search problem in the same way as the cubic nonlinearity. Then the lower bound $N_0 = \Omega(N/\log N)$ from the cubic nonlinearity carries over. With a single marked vertex, the cubic-quintic nonlinearity requires a constant number of atoms in an atomic clock to achieve the necessary time-measurement precision, and this yields the same bound. Thus  the strongest lower bound on the number of particles is the same as for the cubic nonlinearity:
\[ N_0 = \Omega\left( \frac{N}{\log N} \right). \]
To the best of our knowledge, this is the first bound on the number of particles needed for the cubic-quintic Schr\"odinger equation to be a good approximation of the linear, many-body Schr\"odinger equation.

\section{Loglinear Nonlinearity}

\begin{figure}
	\includegraphics{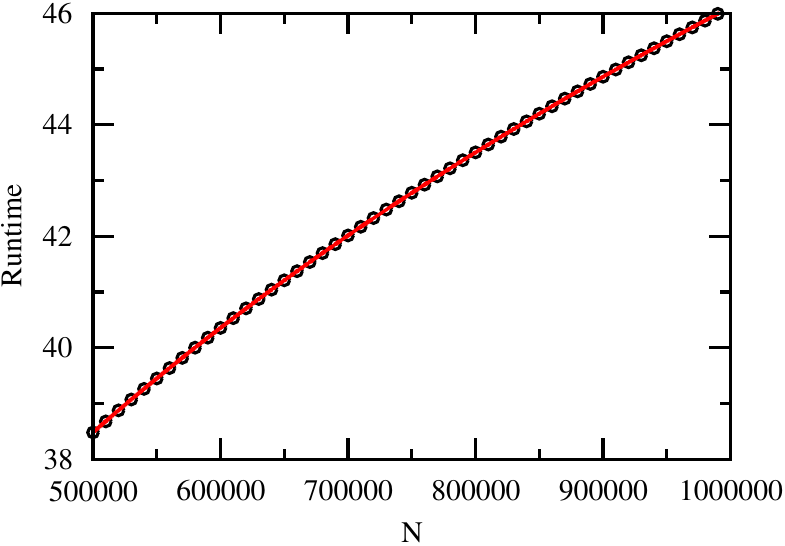}
	\caption{\label{fig:log_runtime_trend}The runtime of search with the loglinear nonlinearity for $k = N^{1/4}$ and $g = N^{1/8}/\log(N/k)$. The black circles were numerically calculated from \eqref{eq:logtime} for $N = 500\,000$ to $N = 1\,000\,000$ with intervals of $10\,000$, and the red solid line is the best-fit curve $t_* = 1.226\,N^{0.261}$.}
\end{figure}

The logarithmic nonlinear Schr\"odinger equation has the form \eqref{eq:NLSE} with $f(p) = \log p$. Then
\[ f_\alpha - f_\beta = \log \left( \frac{N-k}{k} \frac{x}{1-x} \right). \]
Plugging this into \eqref{eq:runtime}, the runtime is given by the integral
\begin{equation}
	\label{eq:logtime}
	t_* = \frac{N}{2\sqrt{k}} \int_{x_0 = k/N}^{1} \frac{1}{1 + g\log \left( \frac{N-k}{k} \frac{x}{1-x} \right)} \sqrt{\frac{1}{(1-x)(Nx-k)}} {\rm d}x.
\end{equation}
Although it's unclear how to directly integrate this, it is possible to bound it. The details are in Appendix \ref{appendix:logruntime}, and it results in
\[ \sqrt{\frac{N}{k}} \frac{1}{g \log \left( \frac{N}{k} \right)} \lesssim t_* \lesssim \sqrt{\frac{N}{k}} \frac{1}{\sqrt{g \log \left( \frac{N}{k} \right)}}, \]
where the notation $f_1(N) \lesssim f_2(N)$ denotes $f_1(N) = O(f_2(N))$, which implies that $f_1(N) \gtrsim f_2(N)$ denotes $f_1(N) = \Omega(f_2(N))$. Numerically, the actual runtime seems to be closer to the lower bound. For example, when $k = N^{1/4}$ and $g = N^{1/8}/\log N$, the regression shown in FIG.~\ref{fig:log_runtime_trend} yields a runtime scaling of $O(N^{0.261})$, whereas the lower bound is $O(N^{1/4})$ and the upper bound is $O(N^{5/16}) = O(N^{0.3125})$. Given the frequent appearance of the ratio $N/k$, we define $R = N/k$. Also, the nonlinearity coefficient $g$ appears with a factor of $\log R$, so we now let $g = O(R^\sigma / \log R)$ rather than $O(N^\kappa)$ from before. Then the bounds are
\[ R^{1/2-\sigma} \lesssim t_* \lesssim R^{1/2-\sigma/2}. \]

For the time-measurement precision, note that $f_\alpha|_{x=1} - f_\beta|_{x=1} = \infty$, so \eqref{eq:width} says the width of the success probability is zero. But from FIG.~\ref{fig:prob_time_mashup}, that can't be right. This incorrect results arises because \eqref{eq:width} was derived by Taylor expanding the success probability about its peak, but for the loglinear nonlinearity, the second derivative at the peak is negative infinity. To get around this, we instead Taylor expand about a nearby point, and the details are worked out in Appendix \ref{appendix:logwidth}. Then the width of the success probability at height $1 - \epsilon$ is bounded by
\[ \Delta t = \Omega\left( \sqrt{\frac{N}{k}} \frac{1}{g \log \left( \frac{N}{k} \frac{1}{\epsilon} \right)} \right). \]
Note that in FIG.~\ref{fig:prob_time_mashup}, we chose $g = O(\sqrt{N}/\log N)$ since $k$ was constant, and it resulted in constant runtimes and widths. So this bound seems tight, and it is further evidence that the runtime $t_*$ is closer to its lower bound.

To achieve this level of time-measurement precision in an atomic clock that utilizes entanglement, we need the number of clock ions scale inversely with $\Delta t$. Also including the $\log N$ qubits to encode the $N$-dimensional Hilbert space, the total ``space'' requirement $S$ is
\[ S = O \left( R^{\sigma - 1/2} + \log N \right). \]
Then the total resource requirement when $\sigma \ge 1/2$ is
\[ \left( 1 + R^{1/2-\sigma} \log N \right) \lesssim ST \lesssim \left( R^{\sigma/2} + R^{1/2-\sigma/2} \log N \right). \]
This is minimized when $\sigma = 1/2$, yielding
\[ \log N \lesssim ST \lesssim R^{1/4} \log N. \]
The upper bound equals the cubic nonlinearity's total resource requirement. So the loglinear nonlinearity is at least as good as the cubic nonlinearity in reducing the time-space resources. Given the numerical results from FIG.~\ref{fig:log_runtime_trend}, the actual total resources seem closer to the lower bound.

Of course, there must be additional resources such that the product of the space requirements and the square of the time requirements is lower bounded by $N$ \cite{Zalka1999, MeyerWong2013}. If the physical system (\textit{e.g.}, a Bose liquid) has $N_0$ particles, then each particle can be at any of the $N$ vertices of the graph, which requires $\log N$ qubits for each particle. Then the ``space'' requirement $S$ is $N_0 \log N$ plus the number of clock ions to achieve the necessary time-measurement precision. That is,
\[ S = O\left( N_0 \log N + R^{\sigma - 1/2} \right) \]
for large $N$. Then
\[ R^{1-2\sigma} (N_0 \log N + R^{\sigma - 1/2}) \lesssim ST^2 \lesssim R^{1-\sigma} (N_0 \log N + R^{\sigma-1/2}). \]
Since this must be lower bounded by $N$,
\[ R^{1-2\sigma} (N_0 \log N + R^{\sigma - 1/2}) = \Omega(N). \]
When $\sigma \le 1/2$, this bound is satisfied regardless of $N_0$. When $\sigma > 1/2$, then
\[ N_0 = \Omega\left( \frac{N R^{2\sigma - 1}}{\log N} \right). \]
As $\sigma$ increases, this bound also increases. But there is no reason to increase $\sigma$ beyond $1/2$, at which $N_0 = \Omega( N \log N )$, because that gives the optimal product of space and time when ignoring $N_0$, and numerically gives constant runtime. So we've given a quantum information-theoretic bound for the number of particles needed for the logarithmic nonlinear Schr\"odinger equation to describe the physical system (\textit{e.g.}, the number of atoms in a Bose liquid), and to the best of our knowledge, it is the first such result.

\section{Conclusion}

Our results indicate that a host of physically realistic nonlinear quantum systems of the form \eqref{eq:NLSE} can be used to perform continuous-time computation faster than (linear) quantum computation. In particular, we've quantified this speedup by analyzing the quantum search problem, and the particular choice of nonlinearity gives rise to different runtimes, requires different levels of time-measurement precision, and necessitates a different number of particles for the nonlinearity to be an asymptotic description of the many-body quantum dynamics. 

\begin{acknowledgments}
This work was partially supported by the Defense Advanced Research Projects Agency as part of the Quantum Entanglement Science and Technology program under grant N66001-09-1-2025, and by the Air Force Office of Scientific Research as part of the Transformational Computing in Aerospace Science and Engineering Initiative under grant FA9550-12-1-0046.
\end{acknowledgments}


\appendix

\section{\label{appendix:cubicquintic}Proof of Runtime Scaling for Cubic-Quintic Nonlinearity}

Recall we are finding the scaling of the runtime
\[ t_* = \frac{\pi}{2} \frac{N k^2(N-k)^2}{2\sqrt{k}} \frac{\sqrt{2}}{\sqrt{\Sigma } \sqrt{\Delta }} \left[ \frac{2 a+b+\sqrt{\Delta}}{\sqrt{\xi +\sqrt{\Delta} (k-N)} } + \frac{-2 a-b+\sqrt{\Delta}}{\sqrt{\xi -\sqrt{\Delta}(k-N)}} \right], \]
where
\[ a = -g N (N-2k) \]
\[ b = gk(N^2-kN-2k) \]
\[ c = -gk^2(N-k-1) + k^2(N-k)^2 \]
and
\[ \Delta = b^2-4a c \]
\[ \Sigma = a+b+c \]
\[ \xi = 2ak + 2cN + b(k+N) \]
when $g = O(N^\kappa)$ and $k = O(N^\lambda)$ (with $0 \le \lambda \le 1$). Let's find the scaling of the individual terms and put them together until we have $t_*$. We have
\[ a = O\left( N^{\kappa + 2} \right) \]
\[ b = O\left( N^{\kappa + \lambda + 2} \right) \]
\[ c = \begin{cases}
	O\left( N^{\kappa + 2\lambda + 1} \right), & \kappa \ge 1 \\
	O\left( N^{2\lambda + 2} \right), & \kappa < 1
\end{cases}. \]
Then
\[ \Delta = \begin{cases}
	O\left( N^{2\kappa + 2\lambda + 4} \right), & \kappa \ge 0 \\
	O\left( N^{\kappa + 2\lambda + 4} \right), & \kappa < 0
\end{cases} \]
\[ \Sigma = \begin{cases}
	O\left( N^{\kappa + \lambda + 2} \right), & \kappa \ge 1 \\
	O\left( N^{\kappa + \lambda + 2} \right), & \kappa < 1, \lambda \le \kappa \\
	O\left( N^{2\lambda + 2} \right), & \kappa < 1, \lambda > \kappa 
\end{cases} \]
\[ \xi = \begin{cases}
	O\left( N^{\kappa + \lambda + 3} \right), & \kappa \ge 1 \\
	O\left( N^{\kappa + \lambda + 3} \right), & \kappa < 1, \lambda \le \kappa \\
	O\left( N^{2\lambda + 3} \right), & \kappa < 1, \lambda > \kappa 
\end{cases} \]
We also have
\[ 2a + b + \sqrt{\Delta} = \begin{cases}
	O\left( N^{\kappa + \lambda + 2} \right), & \kappa \ge 0 \\
	O\left( N^{\kappa/2 + \lambda + 2} \right) & \kappa < 0
\end{cases} \]
This is different, however, from
\[ -2a - b + \sqrt{\Delta} = \begin{cases} 
	O\left( N^{\kappa + 2} \right), & \kappa \ge 1 \\ 
	O\left( N^{\kappa + 2} \right), & 0 \le \kappa < 1, \lambda \le \kappa \\ 
	O\left( N^{\lambda + 2} \right), & 0 \le \kappa < 1, \lambda > \kappa \\
	O\left( N^{\kappa/2 + \lambda + 2} \right) & \kappa < 0
\end{cases} \]
because when $\kappa \ge 0$, the dominant term in $\sqrt{\Delta}$ is $b$, which cancels with $-b$.
The expression
\[ \xi + \sqrt{\Delta} (k-N) \]
is a little tricky. The dominant terms of $\xi$ and $\sqrt{\Delta}(k-N)$ cancel in certain cases. That is, when $\kappa \ge 1$ or $\kappa < 1$ and $\lambda \le \kappa$, then $\xi = 2ak + 2cN + b(k+N)$ is dominated by the $bN$ term. When $\kappa \ge 0$, $\Delta = b^2 - 4ac$ is dominated by the $b^2$ term, so $\sqrt{\Delta}(k-N)$ is dominated by $-bN$. So in these regions, the $bN$'s cancel out, and we should ignore it when computing $\xi + \sqrt{\Delta}(k-N)$, thereby making $\xi = O(2ak + 2cN + bk)$ and $\sqrt{\Delta}(k-N) = O(bk - \frac{2ac}{b}(k-N))$. If we add them together, we get
\[ \xi + \sqrt{\Delta} (k-N) = 2ak + 2cN + 2bk - \frac{2ac}{b}(k-N). \]
Note that $2ak + 2cN + 2bk$ is dominated by $-2gkN^2 + 2k^2N^3$, and $-\frac{2ac}{b}(k-N)$ is dominated by $2gkN^2 - 2kN^3$. Adding these, the $2gkN^2$ factors cancel, leaving $\xi + \sqrt{\Delta} (k-N)$ dominated by $2k^2N^3$. So
\[ \xi + \sqrt{\Delta} (k-N) = O\left( N^{2\lambda + 3} \right). \]
It's easy to see (\textit{i.e.}, we don't have to worry about cancellations) that the scaling is also $N^{2\lambda + 3}$ for other values of $\kappa$ and $\lambda$.
Combining our results,
\[ \frac{2a + b + \sqrt{\Delta}}{\sqrt{\xi + \sqrt{\Delta} (k-N)}} = \begin{cases}
	O\left( N^{\kappa + 1/2} \right), & \kappa \ge 0 \\
	O\left( N^{\kappa/2 + 1/2} \right), & \kappa < 0 \\
\end{cases}. \]
The expression $\xi - \sqrt{\Delta} (k-N)$ is different (easier) because the dominant term no longer cancels. So we have
\[ \xi - \sqrt{\Delta} (k-N) = \begin{cases}
	O\left( N^{\kappa + \lambda + 3} \right), & \kappa \ge 1 \\
	O\left( N^{\kappa + \lambda + 3} \right), & 0 \le \kappa < 1, \lambda \le \kappa \\
	O\left( N^{2\lambda + 3} \right), & 0 \le \kappa < 1, \lambda > \kappa \\
	O\left( N^{2\lambda + 3} \right), & \kappa < 0 \\
\end{cases}. \]
Then
\[ \frac{-2a - b + \sqrt{\Delta}}{\sqrt{\xi - \sqrt{\Delta} (k-N)}} = \begin{cases}
	O\left( N^{\kappa/2 - \lambda/2 + 1/2} \right), & \kappa \ge 1 \\
	O\left( N^{\kappa/2 - \lambda/2 + 1/2} \right), & 0 \le \kappa < 1, \lambda \le \kappa \\
	O\left( N^{1/2} \right), & 0 \le \kappa < 1, \lambda > \kappa \\
	O\left( N^{\kappa/2 + 1/2} \right), & \kappa < 0 \\
\end{cases}. \]
Then
\[ \frac{2a + b + \sqrt{\Delta}}{\sqrt{\xi + \sqrt{\Delta} (k-N)}} + \frac{-2a - b + \sqrt{\Delta}}{\sqrt{\xi - \sqrt{\Delta} (k-N)}} = \begin{cases}
	O\left( N^{\kappa + 1/2} \right), & \kappa \ge 0 \\
	O\left( N^{\kappa/2 + 1/2} \right), & \kappa < 0 \\
\end{cases}. \]
We also have
\[ \sqrt{\Sigma} \sqrt{\Delta} = \begin{cases}
	O\left( N^{3\kappa/2 + 3\lambda/2 + 3} \right), & \kappa \ge 1 \\
	O\left( N^{3\kappa/2 + 3\lambda/2 + 3} \right), & 0 \le \kappa < 1, \lambda \le \kappa \\
	O\left( N^{\kappa + 2\lambda + 3} \right), & 0 \le \kappa < 1, \lambda > \kappa \\
	O\left( N^{\kappa/2 + 2\lambda + 3} \right), & \kappa < 0 \\
\end{cases} \]
Putting all this together,
\[ t_* = \begin{cases}
	O\left( N^{-\kappa/2 + 1/2} \right), & \kappa \ge 1 \\
	O\left( N^{-\kappa/2 + 1/2} \right), & 0 \le \kappa < 1, \lambda \le \kappa \\
	O\left( N^{-\lambda/2 + 1/2} \right), & 0 \le \kappa < 1, \lambda > \kappa \\
	O\left( N^{-\lambda/2 + 1/2} \right), & \kappa < 0
\end{cases} \]
Note that our formula can be reduced to two cases. When $\kappa \ge 1$, then $\lambda \le \kappa$ since $0 \le \lambda \le 1$. Similarly, when $\kappa < 0$, then $\lambda > \kappa$. So we have
\[ t_* = \begin{cases}
	O\left( N^{-\kappa/2 + 1/2} \right), & \lambda \le \kappa \\
	O\left( N^{-\lambda/2 + 1/2} \right), & \lambda > \kappa \\
\end{cases} \]

\section{\label{appendix:logruntime}Bound for Runtime Loglinear Nonlinearity}

\begin{figure}
	\includegraphics{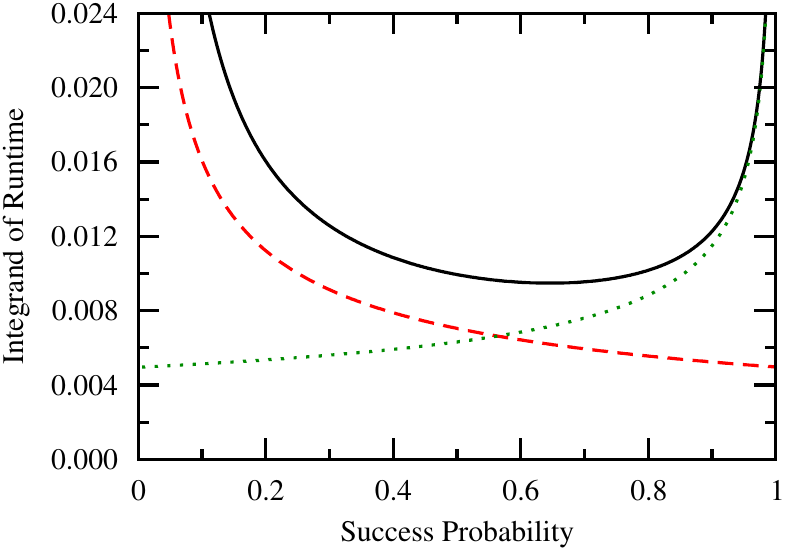}
	\caption{\label{fig:loglowerbound}The integrand of the runtime integral \eqref{eq:logtime} is the solid black curve, and the red dashed and green dotted curves are the integrands of the lower bound integrals \eqref{eq:logtimelowerbound}, all for $N = 1024$, $k = 5$, and $g = 1$.}
\end{figure}

Recall we are finding lower and upper bounds for the runtime
\[ t_* = \frac{N}{2\sqrt{k}} \int_{x_0 = k/N}^{1} \frac{1}{1 + g\log \left( \frac{N-k}{k} \frac{x}{1-x} \right)} \sqrt{\frac{1}{(1-x)(Nx-k)}} {\rm d}x. \]
Let's begin with the lower bound. Splitting the region of integration into two parts, the runtime is bounded below by
\begin{align} 
	\label{eq:logtimelowerbound}
	t_* \ge \frac{N}{2\sqrt{k}} \Bigg[ &\int_{k/N}^{1/2} \frac{1}{1 + g\log \left( \frac{N-k}{k} \frac{1/2}{1-1/2} \right)} \sqrt{\frac{1}{(1-k/N)(Nx-k)}} {\rm d}x \\
	&+ \int_{1/2}^{1} \frac{1}{1 + g\log \left( \frac{N-k}{k} \frac{1}{1-x} \right)} \sqrt{\frac{1}{(1-x)(N\cdot1-k)}} {\rm d}x \Bigg]. \notag
\end{align}
These integrands are shown in FIG.~\ref{fig:loglowerbound} along with the original integrand, illustrating that they are indeed lower bounds. These integrate to
\begin{align*} 
	t_* \ge \frac{N}{2\sqrt{k}} \frac{1}{\sqrt{N-k}} \Bigg[ &\sqrt{\frac{2(N-2k)}{N}} \frac{1}{1 + g \log \left( \frac{N-k}{k} \right)} \\
	&- e^{\frac{1}{2g}} \frac{1}{g} \sqrt{\frac{N-k}{k}} \text{E}_1\!\left( \frac{1 + g \log \left( \frac{2(N-k)}{k} \right)}{2g} \right) \Bigg],
\end{align*}
where $\text{E}_1$ is the exponential integral
\[ \text{E}_1(x) = \int_{x}^\infty \frac{e^{-t}}{t} {\rm d}t, \]
which is bounded by
\[ \frac{1}{2} e^{-x} \log \left( 1 + \frac{2}{x} \right) < \text{E}_1(x) < e^{-x} \log \left( 1 + \frac{1}{x} \right). \]
Then the runtime is lower bounded by
\begin{align*} 
	t_* \ge \frac{N}{2\sqrt{k}} \frac{1}{\sqrt{N-k}} \Bigg[ &\sqrt{\frac{2(N-2k)}{N}} \frac{1}{1 + g \log \left( \frac{N-k}{k} \right)} \\
	&- \frac{1}{\sqrt{2}g} \log \left( 1 + \frac{2g}{1 + g \log \left( \frac{2(N-k)}{k} \right)} \right) \Bigg].
\end{align*}
Now assume that $g = O(N^\kappa)$ with $\kappa > 0$. Then for large $N$, this becomes
\[ t_* = \Omega \left( \sqrt{\frac{N}{k}} \frac{1}{g \log \left( \frac{N}{k} \right)} \right). \]

\begin{figure}
	\includegraphics{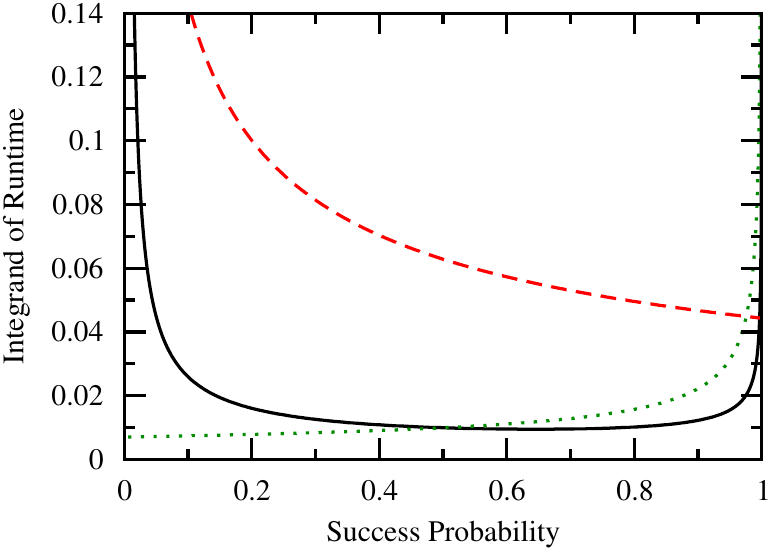}
	\caption{\label{fig:logupperbound}The integrand of the runtime integral \eqref{eq:logtime} is the solid black curve, and the red dashed and green dotted curves are the integrands of the upper bound integrals \eqref{eq:logtimeupperbound}, all for $N = 1024$, $k = 5$, and $g = 1$.}
\end{figure}

Now for the upper bound, we can again split the region of integration into two parts:
\begin{align} 
	\label{eq:logtimeupperbound}
	t_* \le \frac{N}{2\sqrt{k}} \Bigg[ &\int_{k/N}^{1/2} \frac{1}{1 + g\log \left( \frac{N-k}{k} \frac{k/N}{1-k/N} \right)} \sqrt{\frac{1}{(1-1/2)(Nx-k)}} {\rm d}x \\
	&+ \int_{1/2}^{1} \frac{1}{1 + g\log \left( \frac{N-k}{k} \frac{1/2}{1-1/2} \right)} \sqrt{\frac{1}{(1-x)(N/2-k)}} {\rm d}x \Bigg]. \notag
\end{align}
These integrands are shown in FIG.~\ref{fig:logupperbound} along with the original integrand, illustrating that they are indeed upper bounds. The first region, however, is a poor bound, so we expect our result to not be tight. These integrate to
\[ t_* \le \frac{N}{2\sqrt{k}} \Bigg[ \frac{2\sqrt{N-2k}}{N} + \frac{2}{\sqrt{N-2k} \left[ 1 + g \log \left(\frac{N-k}{k}\right) \right]} \Bigg]. \]
Again assuming that $g = O(N^\kappa)$ with $\kappa > 0$ and large $N$,
\[ t_* = O\left( \sqrt{\frac{N}{k}} \right). \]
But this isn't very insightful. It simply says that the nonlinear algorithm is no worse than the linear algorithm. This is expected because our upper bound is not very tight.

\begin{figure}
	\includegraphics{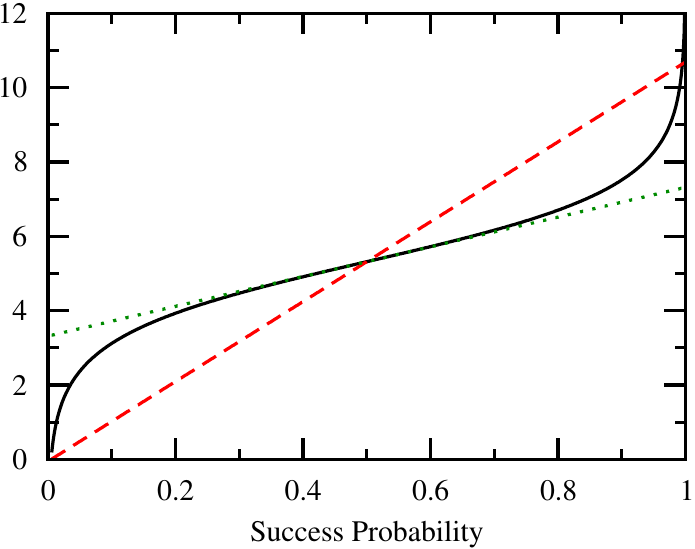}
	\caption{\label{fig:logbetterupperbound}Plot of \eqref{eq:logbetterupperbound} for $N = 1024$ and $k = 5$. The black solid curve is original logarithm, the red dashed curve is the bound from $k/N < x < 1$, and the green dotted curve is the bound from $1/2 < x < 1$.}
\end{figure}

\begin{figure}
	\includegraphics{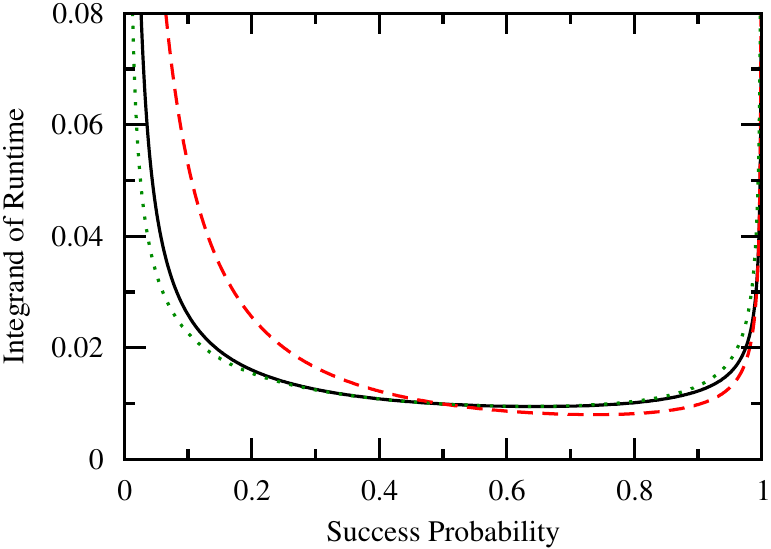}
	\caption{\label{fig:logbetterupperboundintegrand}The integrand of the runtime integral \eqref{eq:logtime} is the solid black curve, and the red dashed and green dotted curves are the integrands of the upper bound integrals \eqref{eq:logtimebetterupperbound}, all for $N = 1024$, $k = 5$, and $g = 1$.}
\end{figure}

To find a tighter bound for the runtime, we instead replace the logarithmic term in the denominator of the runtime integral \eqref{eq:logtime} with a smaller function. In the region $k/N < x < 1/2$, we can use the line connecting those points, and in the region $1/2 < x < 1$, we use the first-order Taylor approximation at $x = 1/2$:
\begin{equation}
	\label{eq:logbetterupperbound}
	\log \left( \frac{N-k}{k} \frac{x}{1-x} \right) \le
	\begin{cases}
		\frac{2}{N-2k} \log \left( \frac{N-k}{k} \right) (Nx-k) & k/N < x < 1/2 \\
		\log \left( \frac{N-k}{k} \right) + 4 \left( x - \frac{1}{2} \right) & 1/2 < x < 1
	\end{cases}.
\end{equation}
The bounds for this logarithm are shown in FIG.~\ref{fig:logbetterupperbound}. Then the runtime is bounded by
\begin{align*}
	\label{eq:logtimebetterupperbound}
	t_*
		&\le \frac{N}{2\sqrt{k}} \int_{k/N}^{1/2} \frac{1}{1 + g\frac{2}{N-2k} \log \left( \frac{N-k}{k} \right) (Nx-k) } \sqrt{\frac{1}{(1-x)(Nx-k)}} {\rm d}x \\
		&\quad+ \frac{N}{2\sqrt{k}} \int_{1/2}^{1} \frac{1}{1 + g \left( \log \left( \frac{N-k}{k} \right) + 4 \left( x - \frac{1}{2} \right) \right)} \sqrt{\frac{1}{(1-x)(Nx-k)}} {\rm d}x.
\end{align*}
These integrands are shown in FIG.~\ref{fig:logbetterupperboundintegrand} along with the original integrand, illustrating that they are indeed upper bounds. They are also tighter than our previous attempt illustrated in FIG.~\ref{fig:logupperbound}. The runtime integrates to
\begin{align*}
	t_* \le \frac{N}{2\sqrt{k}} \Bigg\{
		&\frac{-2 \sqrt{N-2k}}{\sqrt{N} \sqrt{N-2k+2g(N-k)\log\left(\frac{N-k}{k}\right)}} \tan^{-1} \left( \frac{\sqrt{N}}{\sqrt{N-2k+2g(N-k)\log\left(\frac{N-k}{k}\right)}} \right) \\
		&+ \frac{\pi}{\sqrt{N-k}\sqrt{N}} \sqrt{\frac{N^2-3kN+2k^2}{N-2k+2g(N-k)\log\left(\frac{N-k}{k}\right)}} \\
		&+ \frac{2 \tan^{-1} \left( \frac{\sqrt{4gk+N-2gN+gN\log\left(\frac{N-k}{k}\right)}}{\sqrt{N-2k}\sqrt{1+2g+g\log\left(\frac{N-k}{k}\right)}} \right)}{\sqrt{1+2g+g\log\left(\frac{N-k}{k}\right)}\sqrt{4gk+N-2gN+gN\log\left(\frac{N-k}{k}\right)}} \Bigg\}.
\end{align*}
For $g = O(N^\kappa)$ with $\kappa > 0$ and large $N$, this is dominated by the second term and becomes
\[ t_* = O\left( \sqrt{\frac{N}{k}} \frac{1}{\sqrt{g \log \left( \frac{N}{k} \right)}} \right). \]
Combining this with the lower bound, the runtime is bounded between
\[ \sqrt{\frac{N}{k}} \frac{1}{g \log \left( \frac{N}{k} \right)} \lesssim t_* \lesssim \sqrt{\frac{N}{k}} \frac{1}{\sqrt{g \log \left( \frac{N}{k} \right)}}. \]

\section{\label{appendix:logwidth}Width of Success Probability for Loglinear Nonlinearity}

In this appendix, we derive a bound for the width of the sucess probability $x(t)$ by Taylor expanding about $1 - \epsilon$. So we need the first and second derivative, which from \eqref{eq:dxdt} and \eqref{eq:d2xdt2} are
\[ \frac{{\rm d}x}{{\rm d}t} = \pm \sqrt{ \frac{4 k (1-x) (Nx-k) \left[1 + g \left( f_\alpha - f_\beta \right) \right]^2}{N^2} } \]
and 
\[ \frac{{\rm d}^2x}{{\rm d}t^2} = \frac{N}{1 + g(f_\alpha- f_\beta)} \frac{g}{N} \left[ \frac{1}{k} f'_\alpha + \frac{1}{N-k} f'_\beta \right] \left( \frac{{\rm d}x}{{\rm d}t} \right)^2 + 2 \left( \frac{1+g(f_\alpha-f_\beta)}{N} \right)^2 k \left( N - 2Nx + k \right). \]
For the loglinear nonlinearity
\[ f_\alpha - f_\beta = \log \left( \frac{N-k}{k} \frac{x}{1-x} \right) \]
and
\[ \frac{1}{k} f'_\alpha + \frac{1}{N-k} f'_\beta = \frac{1}{x} - \frac{1}{1-x} = \frac{1}{x(1-x)}. \]
So near $x = 1-\epsilon$ for small $\epsilon$ and large $N$,
\[ \left. f_\alpha - f_\beta \right|_{x=1-\epsilon} \approx \log \left( \frac{N}{k} \frac{1}{\epsilon} \right). \]
Still for large $N$, the first derivative is
\[ \left. \frac{{\rm d}x}{{\rm d}t} \right|_{x=1-\epsilon} \approx \pm \sqrt{\frac{k\epsilon}{N}} g \log \left( \frac{N}{k} \frac{1}{\epsilon} \right) \]
and the second derivative is
\[ \left. \frac{{\rm d}^2x}{{\rm d}t^2} \right|_{x=1-\epsilon} \approx - \frac{k g^2 \log^2 \left( \frac{N}{k} \frac{1}{\epsilon} \right)}{N} \]
Then the Taylor expansion is
\[ x(t) \approx (1-\epsilon) + \sqrt{\frac{k\epsilon}{N}} g \log \left( \frac{N}{k} \frac{1}{\epsilon} \right) (t - t_{1-\epsilon}) - \frac{1}{2} \frac{k g^2 \log^2 \left( \frac{N}{k} \frac{1}{\epsilon} \right)}{N} (t - t_{1-\epsilon})^2, \]
where $t_{1-\epsilon}$ is the time in which the success probability is $1 - \epsilon$. Now let's consider the time in which the success probability reaches a height of $1 - \epsilon/2$, which is closer to the peak of $1$. For small $\epsilon$, the first-derivative of $x(t)$ in this region is decreasing towards $0$ because the success probability is approaching the peak (where its derivative is zero). That is, for small $\epsilon$, we're considering the region after the success probability's inflection point. Then the width $\delta t = t_{1-\epsilon/2} - t_{1-\epsilon}$ is a lower bound for the width $\Delta t = t_* - t_{1-\epsilon/2}$, where $t_*$ is the time when the success probability is $1$ (\textit{i.e.}, the runtime). Then the Taylor expansion becomes
\[ 1 - \frac{\epsilon}{2} \approx (1-\epsilon) + \sqrt{\frac{k\epsilon}{N}} g \log \left( \frac{N}{k} \frac{1}{\epsilon} \right) (\delta t) - \frac{1}{2} \frac{k g^2 \log^2 \left( \frac{N}{k} \frac{1}{\epsilon} \right)}{N} (\delta t)^2. \]
This is a quadratic for $\delta t$. Solving it and keeping the highest order terms,
\[ \delta t \approx \frac{\sqrt{\frac{k\epsilon}{N}} g \log \left( \frac{N}{k} \frac{1}{\epsilon} \right)}{\frac{k}{N} g^2 \log^2 \left( \frac{N}{k} \frac{1}{\epsilon} \right)} = \sqrt{\frac{N}{k}} \frac{1}{g \log \left( \frac{N}{k} \frac{1}{\epsilon} \right)}. \]
So the width is bounded by
\[ \Delta t = \Omega \left( \sqrt{\frac{N}{k}} \frac{1}{g \log \left( \frac{N}{k} \frac{1}{\epsilon} \right)} \right). \]


\bibliography{refs}

\end{document}